\documentclass[twocolumn,prd,aps,superscriptaddress]{revtex4}

\usepackage{graphicx}
\usepackage{dcolumn}
\usepackage{bm}

\begin{document}

\title{Dissociation cross sections of ground-state and\\ 
excited charmonia with light mesons in the quark model}

\author{T. Barnes}
\affiliation{Physics Division, Oak Ridge National Laboratory, Oak Ridge,
TN 37831}
\affiliation{Department of Physics and Astronomy, 
University of Tennessee, Knoxville, TN
37996}

\author{E. S. Swanson}
\affiliation{Department of Physics and Astronomy, University of Pittsburgh, 
Pittsburgh, PA 15260} 
\affiliation{Thomas Jefferson National Accelerator Facility, Newport News, 
VA 23606}

\author{C.-Y. Wong}
\affiliation{Physics Division, Oak Ridge National Laboratory, 
Oak Ridge, TN 37831}

\author{X.-M.  Xu}
\affiliation{Physics Division, Oak Ridge National Laboratory, 
Oak Ridge, TN 37831}

\begin{abstract}
\vskip .3 true cm
We present numerical results for the dissociation cross sections 
of ground-state, orbitally- and radially-excited charmonia 
in collisions with
light mesons. 
Our results are derived using the 
nonrelativistic quark model, so all
parameters are determined by fits to the 
experimental meson spectrum.
Examples of dissociation into both exclusive and inclusive final states
are considered.
The dissociation cross sections of 
several C=(+) charmonia may be
of considerable importance for the study of heavy ion collisions,
since these states are expected to be
produced more copiously than the $J/\psi$. 
The relative importance of the productions of ground-state and 
orbitally-excited
charmed mesons in a pion-charmonium collision
is demonstrated through the $\sqrt {s}$-dependent charmonium
dissociation cross sections.
\end{abstract}

\maketitle

\section{Introduction}

The interactions of mesons in systems containing both light and heavy
quarks have long been of interest to hadron physicists. 
For example, some models predict that the
open-charm 
$D$ and $\bar D$ mesons have sufficiently attractive 
residual strong interactions with nucleons
to form ``charmed nuclei" \cite{Bro90,charm_nuclei}. 
Charmed hadron bound states may exist in other systems as well; 
Novikov
{\it et al.}
long ago speculated 
that the nominal 3$^3$S$_1$ $c\bar c$ state 
$\psi(4040)$ might actually be a quasinuclear ``molecule" bound state 
of a 
$D^*{\bar D}^*$ pair \cite{Nov78}. 
Several quark model studies have shown that $Q^2\bar q^2$ mesons should
exist for sufficiently large heavy-quark mass $m_Q$ (for a recent
review see Richard \cite{Ric00}). In a recent ``pedagogical" application,
the $m_Q\to\infty$ limit of the $Q^2\bar q^2$ heavy-light system  
(the so-called ${\cal BB}$ system) has been used 
as a theoretical laboratory for the study of nuclear forces, and nuclear
potential energy curves have been derived using
the nonrelativistic
quark model \cite{Bar99} and lattice gauge theory (LGT) \cite{Mic99}.

Recently, further interest in the interactions of light- and heavy-quark
mesons has arisen in the context of heavy-ion collisions and the search
for the quark-gluon plasma (QGP). One signature proposed for 
the identification of a QGP \cite{Mat86} 
is the suppression of the rate of formation of the $J/\psi$ 
and other 
$c\bar c$ bound states. The long-ranged linear 
confining potential between a $c\bar c$ pair 
would purportedly be screened by a QGP, so a $c\bar c$ pair produced in
the collision would be more likely to separate than to populate 
bound $c\bar c$ resonances.

Direct experimental confirmation of such a suppression 
can be detected for example through the observation of 
lepton pairs from the decay $J/\psi \to \ell^+\ell^-$.
The simplest interpretation of an observed
$J/\psi \to \ell^+\ell^-$ signal would be to assume that 
all $J/\psi$ mesons survive until they decay outside
the interaction region. However, if dissociation reactions such as
$\pi + J/\psi \to D^*\bar D$ 
and
$\rho + J/\psi \to D\bar D$ 
are important, the interpretation of the
experiment is more complicated; a weak $J/\psi$ signal might simply be due
to dissociation through such ``comover absorption" processes.
The actual size of these low-energy charmonium dissociation 
cross sections is currently very controversial,
and their evaluation 
is the subject of this paper. 

\section{Approaches}

A very wide range 
of theoretical estimates of
low-energy charmonium dissociation cross sections
has been reported in the literature, largely 
due to different assumptions for the dominant
scattering mechanisms. We will briefly review 
the three main approaches 
used before presenting new results from our 
quark model calculations.

\subsection{Quark Interchange}

Quark model calculations of charmonium - light hadron
cross sections were first reported by Martins, Blaschke and Quack 
\cite{Mar95}, who used
a quark-interchange model \cite{Bar92} to treat
$\pi + J/\psi$ collisions. 
This reference used standard quark model one gluon exchange (OGE) 
forces (spin-spin
hyperfine and color Coulomb), augmented by a color-independent
confining force that was assumed to act only between
$q\bar q$ pairs (no $qq$ or $\bar q \bar q$ anti-confining interaction).
Since the color Coulomb terms experience
destructive interference between diagrams
(due to color factors) and the large mass of the charm quark makes the hyperfine
term rather weak, Martins {\it et al.} concluded that OGE forces alone
give rather small cross sections; they estimated a cross section for
$\pi + J/\psi \to D^*\bar D + h.c. + D^*\bar D^* $ of $\approx 0.3$~mb
at $\sqrt{s}\approx 4.2$~GeV. However their 
color-independent, $q\bar q$-only confining 
interaction had no color factor cancellation, and so gave a much larger
peak cross section of $\approx 7$~mb near 4.1~GeV. 
These two exclusive processes ($D^*\bar D + h.c.$ and $D^*\bar D^*$)
were found to peak quite close to threshold, and then fell rapidly with 
increasing invariant mass due to suppression from
the Gaussian meson wavefunctions assumed.
(See Fig.2 of Ref.\cite{Mar95}.) 
Subsequently, Wong {\it et al.} \cite{Won00} applied the same approach
to this problem, {\it albeit} with the conventional
quark model $\lambda\cdot\lambda$
dependence for all terms in the interquark Hamiltonian, including the
linear confining potential. The interaction terms assumed were again 
color Coulomb, spin-spin hyperfine and linear scalar confinement. The 
destructive interference between {\it all} diagrams, 
due to zero-sum color factors,
led to a rather smaller 
$\pi + J/\psi \to D^*\bar D + h.c. + D^*\bar D^*$
cross section, with a rather broad peak of roughly 1~mb near 4.1~GeV. 
Wong {\it et al.} also considered $\pi +\psi'$, 
$\rho + J/\psi$ 
and
$\rho + \psi'$ dissociation, and found that these 
cross sections are much larger than
$\pi + J/\psi$ near threshold, due to more favorable kinematics. The
$\rho+(c\bar c)$ processes are exothermic, and so actually diverge at 
threshold.
Finally, Wong {\it et al.} studied the importance of the 
so-called ``post-prior ambiguity" 
\cite{Sch68}
in these calculations; the use of
exact 
$q\bar q$ Hamiltonian wavefunctions in the present paper
eliminates much of this systematic effect, but an important discrepancy
remains due to the use of relativistic phase space and physical masses.

Shuryak and Teaney \cite{Shu98} gave a
comparable rough estimate of 
$\approx 1.2$~mb 
for a low-energy $\pi + J/\psi$ cross section driven by the nonrelativistic
quark model's spin-spin interaction. Actually the specific process they 
considered,
$\pi + J/\psi \to \eta_c + \rho$, is zero at Born order due to a
vanishing color factor; color was not incorporated in their estimate.

\subsection{Meson Exchange}

Charmonium - light hadron scattering can also take place through
t-channel meson exchange. This mechanism was first discussed by
Matinyan and M\"uller \cite{Mat98}, who were motivated 
to study
$\pi + J/\psi$ inelastic scattering 
by the 
great discrepancy between the ca. 7~mb quark-model result of
Martins {\it et al.} \cite{Mar95} and the very small 
low-energy cross sections found using the
Peskin-Bhanot approach \cite{Kha94}.
In the meson-exchange picture, charmonium dissociation reactions 
proceed through t-channel exchange of charmed mesons such as 
$D$ and $D^*$. Matinyan and M\"uller assumed only $D$ exchange, and
found mb-scale cross sections for 
the two low-energy dissociation processes 
$\pi + J/\psi \to D^*\bar D + h.c.$ and
$\rho + J/\psi \to D^*\bar D^* $. 

This work has since been 
generalized to other t-channel exchanges and
effective meson lagrangians. 
Lin, Ko and Zhang had previously proposed an
SU(4) flavor symmetric 
vector-pseudoscalar 
meson effective lagrangian which they had applied to
open-charm meson scattering \cite{Lin99}. Application of this
same lagrangian to the $\pi + J/\psi$ dissociation reaction
$\pi + J/\psi \to D^*\bar D + h.c.$ 
gave a rather large cross section
of $\approx 20-30$~mb for $\sqrt{s}=4-5$~GeV \cite{Lin00};
this was much larger than the 
$D$-exchange 
results of
Matinyan and
M\"uller, due to new three- and four-meson vertices 
in their effective lagrangian.
Haglin \cite{Hag99} introduced a similar SU(4) symmetric
meson lagrangian, and also found rather large 
cross sections of 5-10 mb for $\sqrt{s}=4$-$6$~GeV
for many low energy charmonium dissociation reactions 
(see for example Fig.2 of Ref.\cite{Hag99}). 
Subsequent work 
by Haglin and Gale \cite{Hag01}
showed that the $\pi + J/\psi$ total inelastic
cross section would reach an extremely large value of roughly 100 mb at
$\sqrt{s}=5$~GeV, 
and 
$\rho + J/\psi$ a fantastic 
$\approx 300$~mb (with both still increasing)
in this model, 
assuming pointlike hadron
vertices. 
Similar large 
$\pi + J/\psi$ and
$\rho + J/\psi$ 
cross sections 
have been reported by
Oh, Song and Lee 
in pointlike meson exchange models 
\cite{Oh01}. 

Navarra {\it et al.} \cite{Nav01} have recently questioned the assumption of
flavor SU(4) symmetry; keeping only isospin symmetry, they find
rather smaller cross sections for 
$\pi + J/\psi \to D^*\bar D + h.c.$,
ca. 20-25~mb at $\sqrt{s}=5$~GeV. They confirm that
$D^*$ exchange is much more important numerically in 
meson exchange models than the 
$D$ exchange originally assumed by Matinyan and M\"uller.

Of course it is also incorrect to
assume pointlike hadron form factors. This has been noted
both by 
Lin and Ko \cite{Lin00} and 
Haglin and Gale \cite{Hag01}.
Both collaborations investigated the effect of assuming dipole
forms for the effective three-meson vertices, and 
concluded
that the predicted cross sections were greatly reduced 
(once again to typically 1-10~mb scales) 
with plausible vertex functions. 
(See for example 
Fig.4 of Ref.\cite{Lin00}
and
Fig.7 of Ref.\cite{Hag01}.)
Accurate calculations
of hadronic vertex functions are clearly of 
crucial importance for meson exchange
models of charmonium dissociation. Some results for these form factors, 
obtained from QCD sum rules, 
have been published by Navarra {\it et al.} \cite{Nav00,Dur01}.

\subsection{Diffractive Model}

A high-energy diffractive description of scattering of heavy quarkonia 
which was
developed in 1979 by Peskin and Bhanot \cite{Pes79} has also been applied
to the calculation of charmonium cross sections. It should be stressed that
this method is only justified
at high energies \cite{Pes00}, and then only for deeply-bound
$Q\bar Q$ systems.
It is in essence a gluon-sea model of
high-energy diffractive scattering of 
physically small, high-mass
Coulombic bound states by light hadrons.
This model predicts reasonable 
mb-scale cross sections for 
$J/\psi$ hadronic cross sections at
$\sqrt{s}\geq 10$~GeV
\cite{Pes79,Kha94,Hue00,Arl01}. 
At low energies, however, this mechanism taken in isolation predicts 
extremely small (sub-$\mu$b) cross sections for $J/\psi +\pi$ and
$J/\psi +N$ (see Fig.2 of Ref.\cite{Arl01}). Presumably this means that
the Peskin-Bhanot diffractive scattering
mechanism is unimportant in the low-energy regime
of greatest relevance to QGP searches,
and other mechanisms such as quark interchange and meson exchange
dominate $c\bar c$ strong interactions at these low energies. 
Indeed, a recent comparison \cite{LaSw} with lattice gauge computations 
shows that  
the operator product expansion breaks down for quark masses below
roughly 10 GeV, and therefore the Peskin model is inapplicable to 
light and charmed hadronic physics.

Redlich {\it et al.} \cite{Red00} have argued that these 
diffractive cross sections are actually accurate at low energies,
and if combined with vector dominance can account for the experimental
$\gamma N \to J/\psi + N \to$ open charm  
cross section, whereas the much larger meson-exchange
result for $\sigma(J/\psi + N \to \bar D + \Lambda_c)$ 
of Haglin \cite{Hag99}
greatly exceeds the charm photoproduction measured in
experiment, assuming vector
dominance through the $J/\psi$. 
H\"ufner {\it et al.} \cite{Hue00} however argue 
that this test is misleading,
as the assumption of vector dominance through the $J/\psi$ alone
is unrealistic for these processes. Since the $\psi'$ is 
much closer to open
charm threshold than the $J/\psi$, and
is predicted to 
have larger dissociation cross sections to open charm,
this is clearly a potential source of inaccuracy 
for any $J/\psi$-only vector dominance model.

Finally, QCD sum rule calculations have been performed \cite{Duraes:2002ux}
which find a much larger low-energy cross section than  
the diffraction model Ref.\cite{Kha94},
and are in rough agreement with quark interchange results
near threshold. Although there is approximate agreement of scale 
at low energies, the sum rule results find that the 
exclusive cross sections increase monotonically with energy.
We believe that this is incompatible with hadronic form factors, which
may require the inclusion of higher-dimensional operators in the sum
rule calculations.

\subsection{Synopsis}

Clearly 
the scale of charmonium 
dissociation cross sections at low energies remains an open question. 
Neither the experimental
values nor the dominant scattering mechanisms have been 
convincingly established.
In this currently rather obscure situation we
can best proceed by deriving the predictions of the various
models and searching for the least ambiguous comparisons with experiment,
in as unbiased a manner as possible.
Here we attempt to contribute to this research through a careful
and detailed study of the predictions of one of the theoretical
approaches, the quark interchange model.

\section{Quark Interchange Model}

The Born-order 
quark interchange model approximates hadron-hadron scattering as due 
to a single interaction 
of the standard quark-model interaction Hamiltonian $H_I$
between all constituent pairs in different
hadrons \cite{Bar92}. 
In the current study we specialize to quark interactions that are
simple potentials times spin and color factors,
\begin{equation}
H_I = 
\bigg( 
v_{Cou.}(r)\; {\rm I} + 
v_{conft.}(r)\; {\rm I} + 
v_{ss}(r)\; \vec {\rm S}_i \cdot \vec {\rm S}_j
\bigg)\; 
{T^a \cdot T^a}
\ . 
\end{equation}
The potentials are the standard quark model 
color Coulomb, linear confinement,
and OGE spin-spin hyperfine terms,
$v_{Cou.} = \alpha_s /r$,
$v_{conft.}= -3br/4$,
$v_{ss} = -(8 \pi \alpha_s / 9 m_i m_j )\; \delta_\sigma (\vec r\, )$. 
(The Gaussian-regularized delta function is 
$\delta_\sigma (\vec r\, ) = 
\sigma^3 / \pi^{3/2} \cdot e^{-\sigma^2 r^2 }$.)

Since this Hamiltonian is 
${T^a \cdot T^a}$
in color space,  
quark line rearrangement is required
to give nonzero 
scattering 
amplitudes 
between initial and final color-singlet hadrons
at leading order in $H_I$. 
In the case of $q\bar q$ meson-meson 
scattering, this Born-order amplitude is
given by the sum of the four ``quark Born diagrams" shown in Fig.1.
Each interaction in each diagram has an associated 
``signature" fermion permutation phase,
color factor ${\cal C}$, 
spin matrix element ${\cal S}$,
and a spatial overlap integral ${\cal I}$. The evaluation of these various factors
is discussed in detail in Ref.\cite{Bar92}. 
There are several simplifications in $q\bar q$ meson-meson scattering; the signature
phase is always $(-1)$, 
the flavor factor is diagram independent (and is unity here),
and the color factors are $(-4/9)$ (capture) and $(+4/9)$ (transfer).
The full meson-meson $T$-matrix 
is given by the sum of color Coulomb, linear confinement and OGE spin-spin 
$T$-matrix elements, each of the form
\begin{displaymath}
T_{fi}^{AB\to CD} = 
(-1)\cdot {\cal F} \cdot
\bigg\{ 
(-4/9) 
\cdot \langle {\cal S} \otimes {\cal I}\rangle_{C_1}  +
(-4/9) 
\cdot \langle {\cal S} \otimes {\cal I}\rangle_{C_2} 
\end{displaymath}
\begin{equation}
+\; (+4/9) 
\cdot \langle {\cal S} \otimes {\cal I}\rangle_{T_1}  +
(+4/9) 
\cdot \langle {\cal S} \otimes {\cal I}\rangle_{T_2}  
\bigg\} \ .
\end{equation}
The angle brackets refer to the fact that the spin-
and space-matrix elements do not always factor,
and must in general be evaluated using a Clebsch-Gordon series.
(This complication applies to spin-triplet, orbitally-excited mesons.)
To evaluate the cross section for a given reaction at a given energy,
we first evaluate the overlap integrals (given below)
for each set of orbital magnetic quantum numbers, 
$\langle L_C,L_{Cz}; L_D,L_{Dz} |\;
{\cal I}
\;
|L_A,L_{Az}; L_B,L_{Bz} \rangle$,
using an adaptive  Monte Carlo technique. 
In this method 
we fix $\vec A = A \hat z$, so the overlap integrals are functions of
$\Omega_C$. (The magnitudes of $A$ and $C$ are determined from $\sqrt{s}$ and 
the physical meson masses using relativistic kinematics.)
We then
evaluate spherical harmonic moments 
$c_{lm} = \int d\Omega_C Y_{lm}^*(\Omega_C)\, {\cal I}(\Omega_C)$ of the
overlap integrals, for each diagram and interaction, usually up to $l = 4$.
These spatial overlap integrals
are then combined with the spin matrix elements 
$\langle S_C,S_{Cz}; S_D,S_{Dz} |\;
{\cal S}
\;
|S_A,S_{Az}; S_B,S_{Bz} \rangle$
of ${\rm I}$ and 
$\vec {\rm S}_i \cdot \vec {\rm S}_j$ 
in a Clebsch-Gordon series to form the full T-matrix element
\begin{equation}
T_{fi}^{AB\to CD} = 
\langle J_C,J_{Cz}; J_D,J_{Dz} |\;
T\;
|J_A,J_{Az}; J_B,J_{Bz} \rangle \ .
\end{equation}
Polarized cross sections are then given by
\begin{equation}
\sigma_{fi}^{AB\to CD} = 
{4 E_A E_B E_C E_D \over s}\; {|\vec P_C| \over |\vec P_A|} \;
\int  d\Omega_C \; |T_{fi}^{AB\to CD}|^2
\end{equation}
and the unpolarized cross sections given here are determined by summing over
magnetic quantum numbers as usual.

\begin{figure}
\includegraphics[scale=0.3]{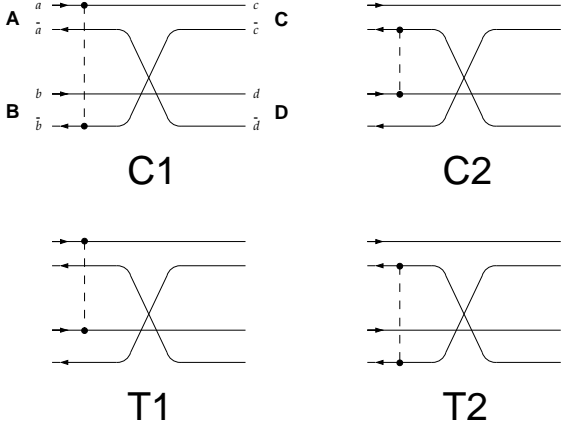}
\caption{\label{fig_1} 
The four quark-interchange meson-meson scattering diagrams
in the ``prior" formalism.}
\end{figure}

Since our $H_I$ consists of simple potentials,
it is convenient to evaluate the
overlap integrals in real space.
(In most recent studies we evaluated these overlap integrals
in momentum space, since they
can be conveniently expressed as convolutions of the
quark-quark momentum-space $T$-matrix 
with external meson wavefunctions; see for example Ref.\cite{Bar01}.
The real-space overlap integrals,
which may be obtained by introducing Fourier transforms in Eqs.(1-4) of
Ref.\cite{Bar01} for the special case $T_{fi}(\vec q, \vec p_1, \vec p_2\, )
= T_{fi}(\vec q\, )$, are

\begin{displaymath}
{\cal I}_{C_1} = 
\int
\hat d^9\! x\;
\psi_A(\vec x_A)\; \psi_B(\vec x_B)\;
\psi_C^*(\vec r\, )\; \psi_D^*(\vec x_A + \vec x_B - \vec r\, )\,
\end{displaymath}
\begin{equation}
v(r) \; \exp \bigg\{ -{i\over 2} \big( \vec A + \mu_C\vec C \big) \cdot \vec x_A 
-{i\over 2} \big( \vec A - \mu_D\vec C \big) \cdot \vec x_B + i \vec A \cdot \vec r \, \bigg\}
\end{equation}

\begin{displaymath}
{\cal I}_{C_2} =  \int
\hat d^9\! x\;
\psi_A(\vec x_A)\; \psi_B(\vec x_B)\; 
\psi_C^*(\vec x_A + \vec x_B - \vec r\, )\; \psi_D^*(\vec r\, )\,
\end{displaymath}
\begin{equation}
v(r) \; \exp \bigg\{ +{i\over 2} \big( \vec A - \mu_C\vec C \big) \cdot \vec x_A 
+{i\over 2} \big( \vec A + \mu_D\vec C \big) \cdot \vec x_B - i \vec A \cdot \vec r \, \bigg\} 
\end{equation}

\begin{displaymath}
{\cal I}_{T_1} =  \int
\hat d^9\! x\;
\psi_A(\vec x_A)\; \psi_B(\vec x_B)\; \psi_C^*(\vec x_B + \vec r\, )\; \psi_D^*(\vec x_A - \vec r\, )\,
\end{displaymath}
\begin{equation}
v(r) \; \exp \bigg\{ -{i\over 2} \big( \vec A + \mu_C\vec C \big) \cdot 
\vec x_A +{i\over 2} \big( \vec A + \mu_D\vec C \big) \cdot \vec x_B 
+ i \vec A \cdot \vec r \, \bigg\}
\end{equation}

\begin{displaymath}
{\cal I}_{T_2} =  \int 
\hat d^9\! x\;
\psi_A(\vec x_A)\; \psi_B(\vec x_B)\; \psi_C^*(\vec x_B + \vec r\, )\;
\psi_D^*(\vec x_A - \vec r\, )\,
\end{displaymath}
\begin{equation}
v(r) \; \exp \bigg\{ +{i\over 2} \big( \vec A - \mu_C\vec C \big) \cdot 
\vec x_A -{i\over 2} \big( \vec A - \mu_D\vec C \big) \cdot \vec x_B 
+ i \vec A \cdot \vec r \, \bigg\}
\end{equation}
Where the measure is  
$\hat d^9\! x\equiv 
d^3r\, d^3x_A\, d^3x_B\, / (2\pi)^3$. 
We also introduced  
$\mu \equiv 2m_q/(m_q + m_{\bar q})$, and 
the identities 
$\mu_A=\mu_B=1$ and
$\mu_C+\mu_D=2$ were used 
in deriving the overlap integrals; these relations are valid for
processes of the type
$(n\bar n)+(c\bar c) \to (n\bar c)+(c\bar n)$.
The spatial wavefunctions above are the usual nonrelativistic 
quark potential model functions 
$\psi(\vec r_{q\bar q})$, normalized to 
$\int d^{\, 3}r\, |\psi(\vec r\, )|^2 = 1$.
The wavefunctions employed in this paper to evaluate these overlap
integrals are numerically determined eigenfunctions of the full 
quark model Hamiltonian (with interaction given by Eq.(1)).

Since we use relativistic phase space and physical masses in evaluating our
cross sections, there is a post-prior ambiguity in our results \cite{Sch68}. 
The overlap integrals given above are the ``prior" form, in which the 
$H_I$ interaction takes place prior to rearrangement (Fig.1). In the ``post" form,
rearrangement followed by interaction, the scattering amplitude is given by a
different set of spin matrix elements and 
overlap integrals. 
The post overlap integrals for the two capture diagrams
are

\begin{displaymath}
{\cal I}_{C_1}^{post} =
\int 
\hat d^9\! x\;
\psi_A(\vec r\, )\;
\psi_B(\vec x_C + \vec x_D - \vec r\, )\,
\psi_C^*(\vec x_C)\;
\psi_D^*(\vec x_D)\;
\end{displaymath}
\begin{equation}
v(r) \; 
\exp
\bigg\{
+{i\over 2} 
\big( 
\vec A + \mu_D\vec C
\big) 
\cdot 
\vec x_C 
-{i\over 2} 
\big( 
\vec A - \mu_D\vec C
\big) 
\cdot 
\vec x_D 
- i \vec C \cdot \vec r \,
\bigg\}
\end{equation}

\begin{displaymath}
{\cal I}_{C_2}^{post} =
\int 
\hat d^9\! x\;
\psi_A(\vec x_C + \vec x_D - \vec r\, )\;
\psi_B(\vec r\, )\,
\psi_C^*(\vec x_C)\;
\psi_D^*(\vec x_D)\;
\end{displaymath}
\begin{equation}
v(r) \; 
\exp
\bigg\{
+{i\over 2} 
\big( 
\vec A - \mu_C\vec C
\big) 
\cdot 
\vec x_C 
-{i\over 2} 
\big( 
\vec A + \mu_C\vec C
\big) 
\cdot 
\vec x_D 
+ i \vec C \cdot \vec r \,
\bigg\}
\end{equation}
The transfer diagrams T1 and T2 (Fig.1) in post and prior
formalisms are identical.

\section{Results}

We have obtained Born-order quark-model results for
{\it i}) dissociation cross sections of ground-state and excited charmonia
into exclusive final states, and
{\it ii}) total inelastic dissociation cross sections from the quark 
interchange mechanism.
These results assume the quark interchange model described 
in the previous section. In the quark model Hamiltonian we assume 
a quark-gluon coupling
constant $\alpha_s =0.6$, 
$\sigma =0.9$~GeV in the spin-spin hyperfine term and
a string tension of $b=0.16$ in the linear confinement term. 
The light and charmed quark
masses are taken to be 
0.33 GeV and 1.6 GeV, respectively. These parameters can reproduce
the $I=2$ $\pi \pi$ $S$-wave experimental phase shifts \cite{Bar01}, 
and are used in 
the following subsections to calculate charmonium dissociation cross sections
which are average values obtained with ``prior'' and ``post'' forms. 
Cross sections for $\pi^+$ scattering are presented in the following sections,
other pion cross sections may be obtained assuming isospin symmetry.

\subsection{Dissociation cross sections of excited charmonia}

The principal mechanism for production of charmonia at small-$x$
in heavy ion collisions
is thought to be the two-gluon fusion process $gg\to c\bar c$. 
One therefore expects $J/\psi$ production to be 
relatively weak, since the formation
of C=($-$) states requires an additional gluon. The
C=(+) mesons that have especially large $gg$ couplings, such as 
$\eta_c$, $\chi_{c0}$ and (to a lesser extent) $\chi_{c2}$ and their,
as yet unidentified, radial excitations
should be the dominant $c\bar c$ states produced. The relative strengths
of $c\bar c$ couplings to glue are dramatically illustrated by the
total widths of charmonia; the 
$\eta_c$ and $\chi_{c0}$ 
total widths, thought to be due mainly to $c\bar c \to gg$,
are two orders of magnitude larger 
than the $J/\psi$ total width. Even the smaller $\chi_{c2}$ width is
roughly $20$ times the $J/\psi$ width.
This suggests that the production of charmonia 
from a quark gluon plasma is probably dominated by
these C=(+) states rather than $J/\psi$, 
so the evolution of C=(+) states produced in 
a heavy-ion collision may be more important for understanding charm
production than the $J/\psi$. The possibility that much of the $J/\psi$
signal originates from radiative transitions of parent $\chi_{cJ}$ states
\cite{Schu,Vogt99} 
also suggests that an understanding of the interactions of these
C=(+) and 
excited $c\bar c$ states with light hadrons may be of great
importance for simulations of heavy flavor production in 
heavy ion collisions.

It is straightforward to determine the
dissociation cross sections of $c\bar c$ states other than the $J/\psi$
in the quark interchange model;
one simply changes the external state attached to each of the 
four scattering diagrams of Fig.1. There is a technical complication 
with spin-triplet, orbitally-excited charmonia, since the
spin and space degrees of freedom do not factor trivially in
these states, unlike the
scattering of S-wave mesons considered previously \cite{Won00}. 
Instead we must evaluate overlap integrals and spin matrix elements for
each set of magnetic quantum numbers, which are then 
combined using the appropriate Clebsch-Gordon coefficients to give scattering
amplitudes of mesons with definite $J$ 
(e.g. $\pi + \chi_{cJ} \to \bar D D^* $).

\subsection{Total $\pi$-charmonium 
dissociation cross sections from constituent interchange}

\begin{figure}
\includegraphics[scale=0.4]{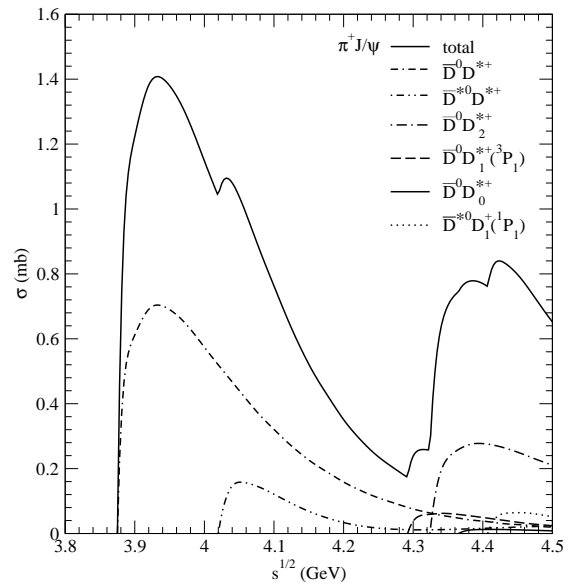}
\caption{\label{fig_pipsi} 
Theoretical $\pi^+ J/\psi$ cross sections in the quark interchange model.
The figure shows all nonzero partial cross sections open to
$\sqrt{s}=4.5$~GeV; the total cross section, obtained by summing these
and their charge conjugate channels, is shown as a solid line.  
}
\end{figure}

\begin{figure}
\includegraphics[scale=0.4]{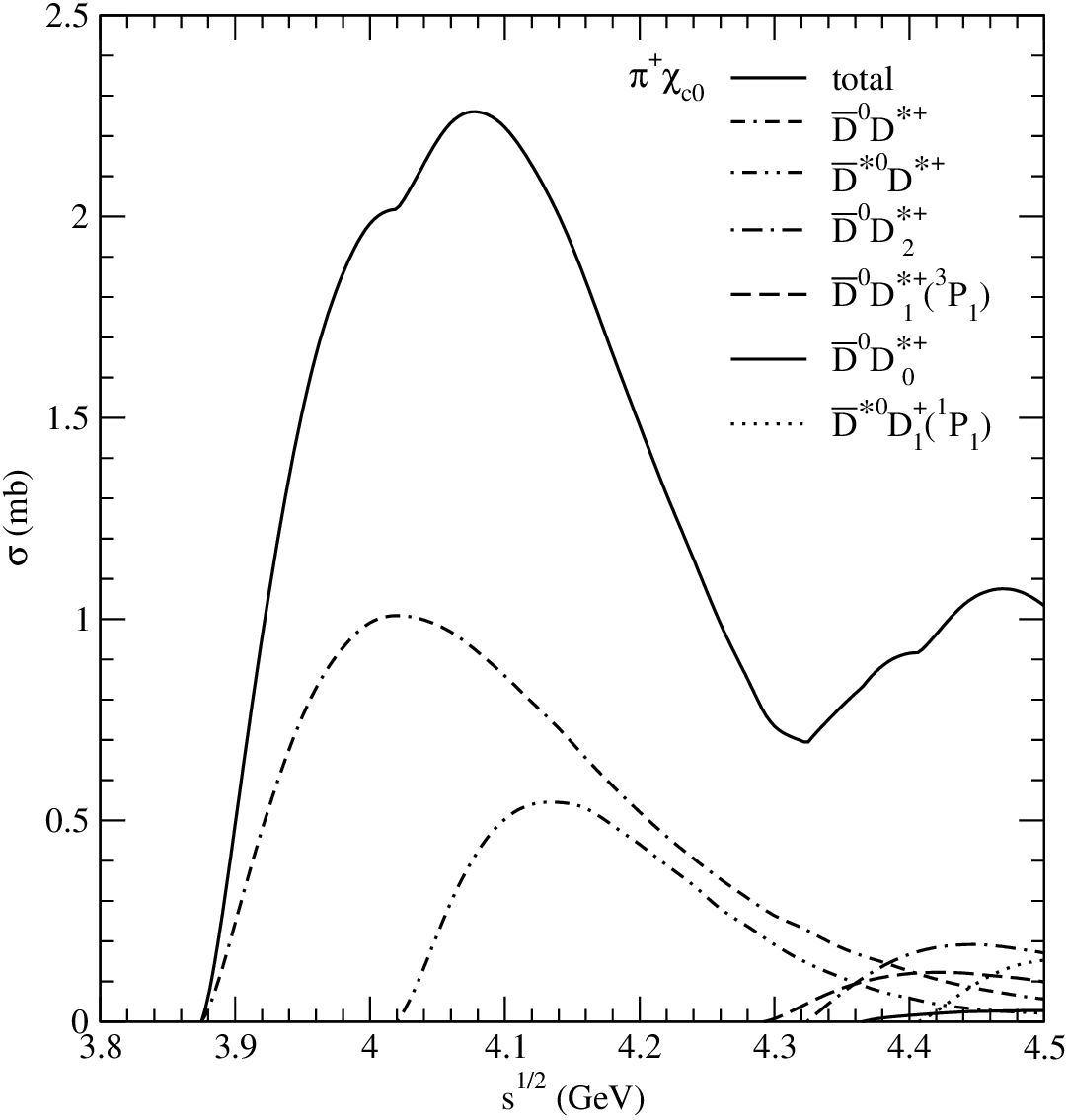}
\caption{\label{fig_pichi0} $\pi^+ \chi_{c0}$ cross sections,
legend as in Fig.2.
}
\end{figure}

\begin{figure}
\includegraphics[scale=0.4]{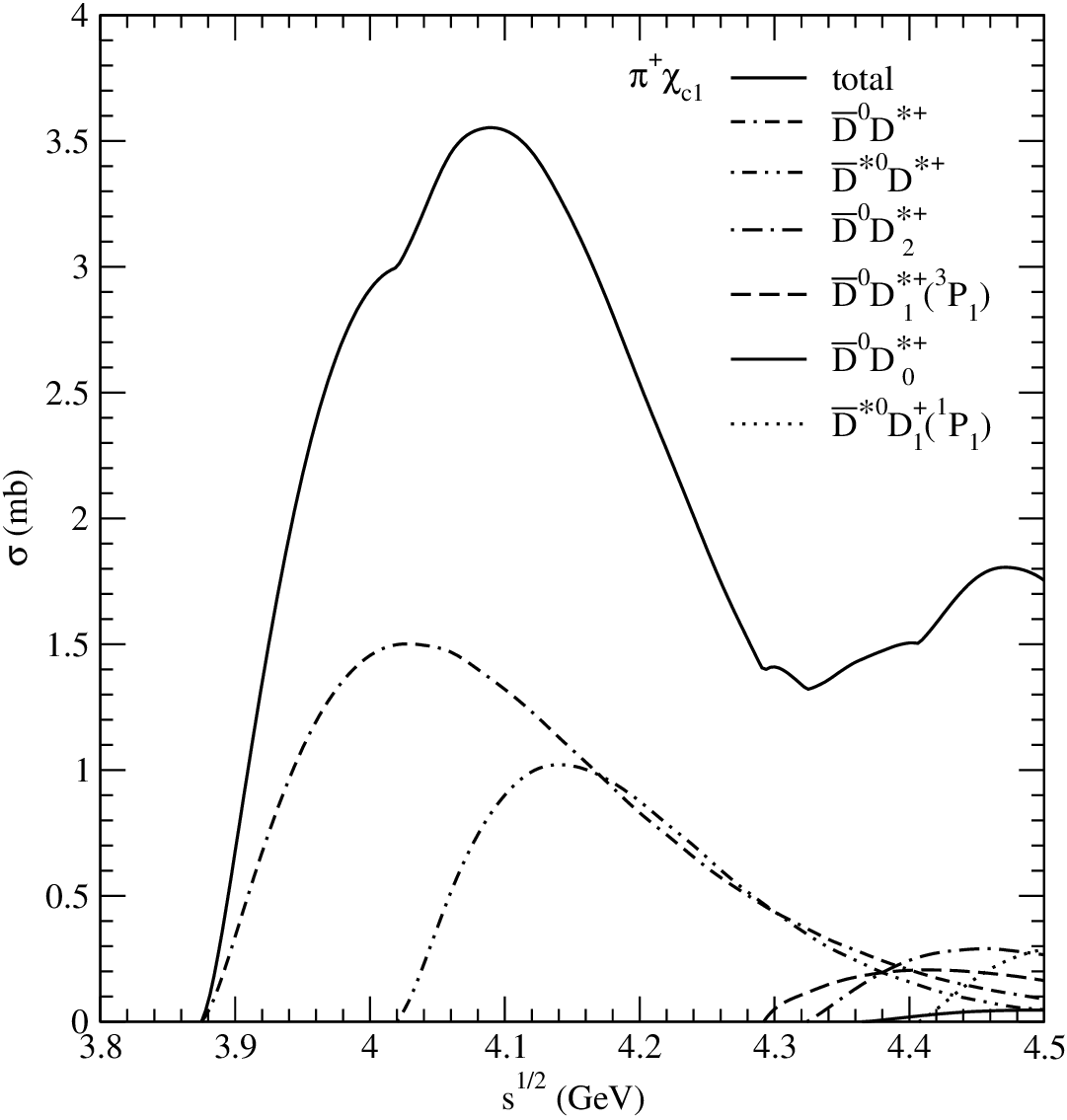}
\caption{\label{fig_pichi1} 
$\pi^+ \chi_{c1}$ cross sections,
legend as in Fig.2.
}
\end{figure}

\begin{figure}
\includegraphics[scale=0.4]{bswx_fig5.eps}
\caption{\label{fig_pichi2}
$\pi^+ \chi_{c2}$ cross sections,
legend as in Fig.2.
}
\end{figure}

\begin{figure}
\includegraphics[scale=0.4]{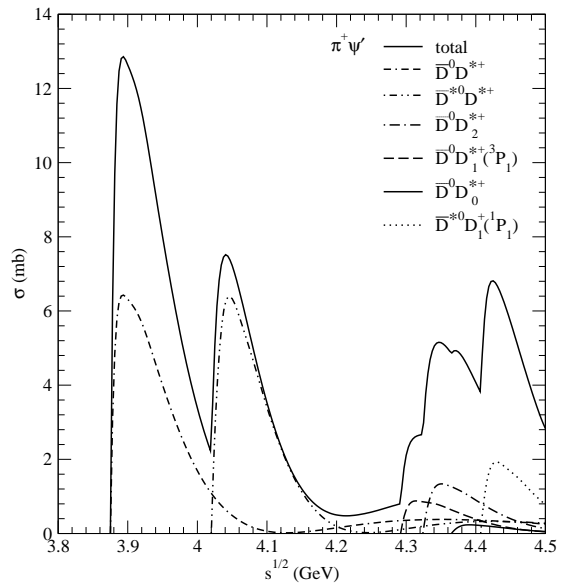}
\caption{\label{fig_pipsip}
$\pi^+ \psi'$ cross sections,
legend as in Fig.2.
}
\end{figure}

\begin{figure}
\includegraphics[scale=0.4]{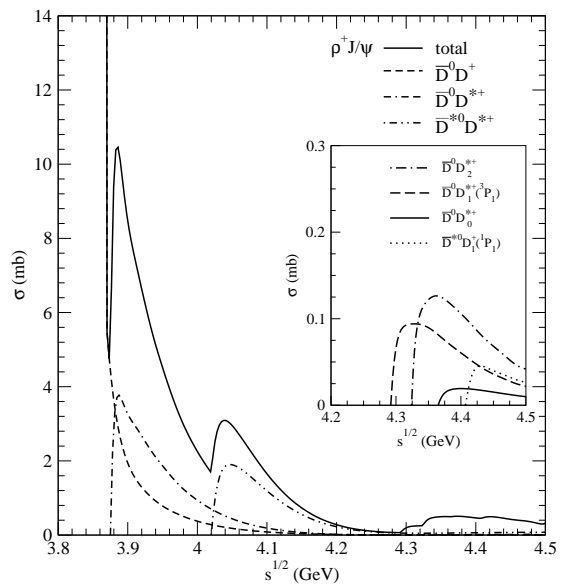}
\caption{\label{fig_rhopsi}
$\rho^+\psi$ cross sections,
legend as in Fig.2.
}
\end{figure}

\begin{figure}
\includegraphics[scale=0.4]{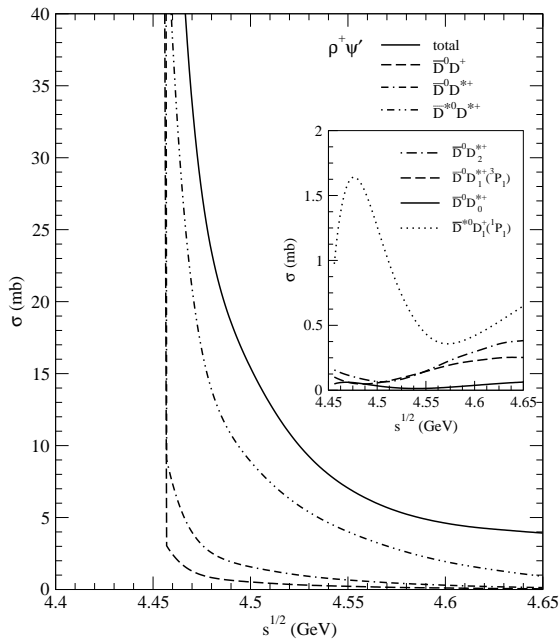}
\caption{\label{fig_rhopsip} 
$\rho^+\psi'$ cross sections,
legend as in Fig.2 but continued to 
$\sqrt{s} = 4.65$~GeV.
}
\end{figure}

We have evaluated the $\pi J/\psi$, $\pi \psi'$, and $\pi \chi_{cJ}$  
exclusive and total cross sections up to 
4.5 GeV in the center of mass
frame. The six final states with nonzero couplings in the model
which are open in this regime
are 

\begin{equation}
\pi^+ + c\bar {c} \to \bar {D}^0 D^{*+} + c.c.
\end{equation}
\begin{equation}
\pi^+ + c\bar {c} \to \bar {D}^{*0} D^{*+}
\end{equation}
\begin{equation}
\pi^+ + c\bar {c} \to \bar {D}^0 D^{*+}_2 + c.c.
\end{equation}
\begin{equation}
\pi^+ + c\bar {c} \to \bar {D}^0 D^{*+}_1 ( ^3P_1) + c.c.
\end{equation}
\begin{equation}
\pi^+ + c\bar {c} \to \bar {D}^0 D^{*+}_0 + c.c.
\end{equation}
\begin{equation}
\pi^+ + c\bar {c} \to \bar {D}^{*0} D^+_1 ( ^1P_1) + c.c.
\end{equation}

The masses of the 
$\bar {D}^0$, $D^{*+}$, $\bar D^{*0}$, $D^+$  and $D_2^{*+}$
mesons are taken from the 2002 PDG compilation 
\cite{Hagi02}. 
The $D_0^{*+}$ vector meson is assumed to 
have a mass of 2.5 GeV. The spin-triplet state  $D_1^{*+} (^3P_1)$ 
and the spin-singlet state  $D_1^+ (^1P_1)$ 
mix to form the observed state
$D_1(2420)^+$ and another unobserved state \cite{God91}.
We assume masses for the 
$D^{*+}_1 (^3P_1)$ and $D^+_1 (^1P_1)$ of  2.427 and 2.4 GeV, respectively.

The total inelastic cross sections to 4.5 GeV are shown in Figs.2-8.
Results for
exclusive reactions are as shown in the figures. The total cross section
also includes charge conjugation final states where appropriate.

We note the following general features of the cross sections. All 
cross sections rise rapidly according to threshold
kinematics and subsequently fall off at a scale of $\Lambda_{QCD}$
as expected for exclusive flavor exchange reactions. The $\psi'$ cross
sections are roughly ten times larger than corresponding $\psi$ cross
sections. This is in accord with the ratio $r_\psi' / r_\psi \approx 2$ and
the notion that cross sections increase with hadron size, 
although we stress
that {\it 
no simple scaling relationship exists in the quark interchange model}.
We note that the ratio of $\psi'$ to $\psi$ is substantially smaller 
than the factor of
5000 predicted by the Peskin-Bhanot computation \cite{Pes79}. The
$\psi'$ cross sections tend to fall more rapidly than those for $\psi$. We
suspect that this is due to the node in the $\psi'$ radial wavefunction
which manifests itself as a zero in the cross section about 200 MeV above
threshold, causing the cross section to drop more rapidly than that for
the ground state. Again, this feature can be expected to be quite general. 
Finally, we note that the $\chi_{cJ}$ cross sections grow with 
angular momentum $J$, which is naively expected due to the increasing 
number of $J_z$ states present.

A simple parameterization of these cross sections may prove useful
for further numerical investigations. We have found that it
is possible to fit many of our numerical cross sections with 
a simple functional form which is motivated by the expected
threshold behavior with an exponential decay representing
suppression due to flavor exchange:

\begin{equation}
\sigma(s) = \sigma_{max} \left( \epsilon\over \epsilon_{max}\right)^{p} 
\exp\left(p(1-\epsilon/\epsilon_{max})\right)
\end{equation}

\noindent
where $\epsilon = \sqrt{s} - M_C - M_D$ and $p = 1/2 + L_{min}^{CD}$
for endothermic reactions and 
$p = -1/2 + L_{min}^{CD}$ for exothermic reactions. Here $L_{min}$ refers
to the minimum possible value for the total 
orbital angular momentum of the final
state consistent with conservation of angular momentum and parity.  One expects
this to dominate the threshold behavior of a given reaction. In practice
we do find that many cross sections are well described by assuming that
the orbital 
angular momentum in the initial channel is zero, however in general many waves
contribute and it is more convenient to simply fit the value of $p$. We have
found that this procedure describes all of our $\pi+c\bar c$ reactions
quite well (however, this is not true for $\rho+c\bar c$). 
Results
for the parameters $p$, $\sigma_{max}$, and $\epsilon_{max}$ 
are presented in the
Table~I. 
The $D^*_1$ and $D_1$ referred to in the column headings represent 
the $D_1(^3P_1)$ and $D_1(^1P_1)$ states respectively. All channels 
except $D^*D^*$  include
charge conjugate reactions in the fit parameters.

\begin{table}
\caption{Cross Section Fit Parameters}
\begin{ruledtabular}
\begin{tabular}{lllllll}
& $DD^*$ & $D^*D^*$ & $DD_2^*$  & $DD_1^*$  & $DD_0^*$  & $D^*D_1$ \\
\hline
$\pi J/\Psi \to $ & & & & & & \\
$p$            &   0.53  & 0.84 & 0.64 & 0.58 & 0.67 & 1.16 \\
$\sigma_{max}$ &   1.40  & 0.154 & 0.562 & 0.012 & 0.026 & 0.127 \\
$\epsilon_{max}$ & 0.059 & 0.044 & 0.074 & 0.050 & 0.050 & 0.052 \\
$\pi \Psi' \to $ & & & & & & \\
$p$            &   0.67  & 1.17 & 0.84 & 0.74 & 0.83 & 1.42 \\ 
$\sigma_{max}$ &   13.04 & 6.30 & 2.66 & 1.78 & 0.466 & 3.81 \\
$\epsilon_{max}$ & 0.027 & 0.034 & 0.033 & 0.032 & 0.032 & 0.037 \\
$\pi \chi_{c0} \to $ & & & & & & \\
$p$            &   1.60  & 1.86  &  1.34 & 1.59 & 0.90 & 2.03 \\
$\sigma_{max}$ &   2.02  & 0.555 & 0.380 & 0.244 & 0.054 & 0.306 \\
$\epsilon_{max}$ & 0.147 & 0.114 & 0.123 & 0.130 & 0.141 & 0.108 \\
$\pi \chi_{c1} \to $ & & & & & & \\
$p$            &   1.63  & 1.84 & 1.44 & 0.95 & 1.57 & 1.93 \\
$\sigma_{max}$ &   3.02  & 1.03 & 0.576 & 0.406 & 0.092 & 0.578 \\
$\epsilon_{max}$ & 0.154 & 0.122 & 0.130 & 0.120 & 0.132 & 0.114 \\
$\pi \chi_{c2} \to $ & & & & & & \\
$p$            &   1.65  & 1.84 & 1.10 & 1.52 & 1.36 & 1.89 \\
$\sigma_{max}$ &   3.64  & 1.37 & 0.724 & 0.598 & 0.122 & 0.790 \\
$\epsilon_{max}$ & 0.157 & 0.127 & 0.132 & 0.138 & 0.133 & 0.116 \\
\end{tabular}
\end{ruledtabular}
\end{table}

\section{Summary and Conclusions}

Total charmonium dissociation cross sections have been computed up
to 4.5 GeV in the center of mass of the $\pi+c\bar c$ system. The
computations employ standard constituent quark model dynamics and
all parameters are fixed by spectroscopic data. Exact 
numerical wavefunctions have been employed in the computations to
minimize post-prior discrepancy.
We have also presented results
for positive charge conjugation $\chi_{cJ}$ dissociation. 
Gluon fusion arguments indicate that these states should be preferentially
produced over negative charge conjugation states in small-$x$ 
heavy ion collisions.

It is of interest to speculate on the high energy limit of these
quark model cross sections. It is apparent from the figures that 
the cross sections we find for channels that open at higher energies
decrease in scale    
as the channel threshold increases. This is expected since the 
higher channels 
have a larger momentum mismatch with the initial state. 
Thus, exchanged quarks must probe the higher momentum region of the initial
hadronic wavefunctions. This implies that the peaks of high mass channels
will be approximately exponentially suppressed in E$_{cm}$, due to wavefunction
suppression of the amplitudes.  
We note that final state hadronic
wavefunctions do not affect this argument, since they are effectively 
averaged over all length scales near threshold.  Furthermore, uncertainties
due to relativistic effects or higher Fock state components will not
change the argument as long as the momentum transfer probes the confinement
region of the wavefunctions.  Uncertainties in the structure of the 
pion arising from nonperturbative spontaneous chiral symmetry
breaking effects are also unimportant here, 
since the well-understood charmonium wavefunctions alone suffice to give
these general cross section features.
Finally, we note that there is an additional suppression 
due to the nodal structure of highly excited wavefunctions.

Once the scattering energy is large enough to probe the Coulombic region
of the hadronic wavefunction we expect to find 
a power-law suppression of the
cross section peak rather than an exponential suppression. This power law is
further weakened by the nodal suppression mentioned above.  We stress
that exclusive cross section peaks must still fall rapidly, even in this
perturbative regime.

For two-to-two scattering the
behavior of inclusive cross sections depends 
on the general behavior
noted above and the density of states, which gives the rate at which 
new channels open with increasing $\sqrt{s}$. Quark models
suggest that the density of $q\bar q$ resonances grows as a power of mass,
thus the total dissociation cross sections must decrease roughly exponentially
while in the confinement regime, and thereafter as a power in the 
perturbative regime. At very high energies the gluonic flux tube may be excited,
leading to an exponential increase in the density of states \cite{H} since the
flux tube contains infinitely many degrees of freedom. However, wavefunction 
suppression will again be exponential due to severe suppression of amplitudes
containing multiply-excited string modes with 
ground state string configurations.
Thus, for the case of two-body to two-body inclusive flavor-exchange 
scattering, 
nonperturbative effects
cannot be ignored, and the cross section should decrease with 
increasing center of mass 
energy.

The constituent quark model provides a microscopic foundation
for the exploration of hadronic interactions at low energy. Thus
all relevant reactions may be computed with the addition of no new
parameters. This stands in contrast to effective models which must
introduce new couplings and form factors, and which suffer from 
confusion over the correct degrees of freedom and dynamics to be
employed (there are no obvious symmetries to guide the construction
of effective lagrangians in this energy regime). Sum rule and 
pQCD computations similarly suffer from the notoriously poor convergence
properties of QCD (as exemplified in renormalon ambiguities) and from
the difficulty in extracting observables from condensates. We regard
the constituent quark model as the most reliable tool for the investigation
of these issues and in future plan to apply it to charmonium-nucleon
scattering and other reactions of interest to RHIC.

\begin{acknowledgments}

We would like to thank our colleagues in the 
PHENIX Collaboration and at the 
University of Rostock for many discussions of
heavy ion collision physics, and B.M\"uller for originally suggesting 
this as a research topic. ESS thanks D.Boyanovsky for discussions 
of theoretical descriptions of high energy cross sections.
The work of Barnes, Wong and Xu was supported in part by the 
Division of Nuclear Physics, 
U. S. Department of Energy, and by the Laboratory Directed Research and
Development Program at
Oak Ridge National Laboratory, 
under Contract No. DE-AC05-00OR22725 managed by 
UT-Battelle, LLC. 
Swanson was supported by the U.S. Department of Energy under contracts
DE-FG02-00ER41135  and  DE-AC05-84ER40150 under which the Southeastern
Universities Research Association operates the Thomas Jefferson 
National Laboratory.
X.-M. Xu thanks the nuclear theory group and PHENIX group at ORNL for their
kind hospitality, and acknowledges additional 
support from the CAS Knowledge
Innovation Project No. KJCX2-SW-N02 and the 
National Natural Science Foundation of China under Grant No. 10135030.
\end{acknowledgments}

\end{document}